%% file: final.tex
\documentclass[12pt]{article}
\usepackage{graphics}
\usepackage{epsfig}
\usepackage{latexsym}
\usepackage{colordvi}
\usepackage{amsmath}
\usepackage{amssymb}
\usepackage{cite}

\def\simkl{^<\hskip -2.5mm_\sim}
\newcommand{\fun}{$F_2^{\gamma}(x,Q^2)$ }

\newcommand{\gam}{^{\gamma}}
\newcommand{\fund}{$F_2^{\gamma}(x,Q^2)$}
\newcommand{\be}{\begin{equation}}
\newcommand{\ba}{\begin{eqnarray}}
\newcommand{\ea}{\end{eqnarray}}
\newcommand{\etal}{{\it et al.}}
\newlength{\figwidth}
\newlength{\figheight}
\def\z0{Z}

\def\ie{{i.e.} }

\title{
\begin{flushright}
\begin{small}
IFT-2003-30 \\
\end{small}
\end{flushright}
\vskip 1.5cm
CJK-Improved 5 Flavour LO Parton Distributions in the Real Photon
}

\author{
F.~Cornet $^a$, P.~Jankowski $^b$, M.~Krawczyk $^b$\\
{\small\it
$^a$ Departamento de F\'{\i}sica Te\'orica y del Cosmos, Universidad de 
Granada},\\
{\small\it Campus de Fuente Nueva, E-18071, Granada, Spain}\\
\vspace{0.1cm}
{\small\it $^b$ Institute of Theoretical Physics, Warsaw University, 
ul. Ho\.za 69, 00-681 Warsaw, Poland}\\
}

\begin{document}

\maketitle

\begin{abstract}

Radiatively generated, LO quark ($u,d,s,c,b$) and gluon densities in the real,
unpolarized photon, improved in respect to our previous paper \cite{cjkl}, are
presented. We perform three global fits to the $F_2\gam$ data, using the LO 
DGLAP evolution equation. We improve the treatment of the strong coupling 
running and used lower values of $\Lambda_{QCD}$, as we have found that the too 
high values adopted in the previous work caused the high $\chi^2$ of the fits. 
In addition to the modified FFNS$_{CJKL}$ model, referred to as FFNS$_{CJK}$1 we
analyse a FFNS$_{CJK}$2 model in which we take into account the resolved-photon 
heavy-quark contribution. New CJK model with an improved high-$x$ behaviour of 
the \fun is proposed. Finally, in the case of the CJK model we abandon the 
valence sum rule imposed on the VMD input densities. New fits give $\chi^2$ per 
degree of freedom about 0.25 better than the old results. All features of the 
CJKL model, such as the realistic heavy-quark distributions, good description of
the LEP data on the $Q^2$ dependence of the $F_2\gam$ and on $F_{2,c}\gam$ are 
preserved. Moreover we present results of an analysis of the uncertainties of 
the CJK parton distributions due to the experimental errors. It is based on the
Hessian method used for the proton and very recently applied for the photon
by one of us. Parton and structure function parametrizations of the best fits in
both FFNS$_{CJK}$ and CJK approaches are made accessible. For the CJK model we 
provide also sets of test parametrizations which allow for calculation of 
uncertainties of any physical value depending on the real photon parton 
densities.
\end{abstract}

\clearpage


\section{Introduction}

In this paper we continue our recent analysis of the LO unpolarized real photon
parton distributions, \cite{cjkl}, improving and broadening our research. The 
main topic of our previous work was the description, within the DGLAP 
evolution framework, of the heavy, charm- and bottom-, quark contributions to 
the photon structure-function, \fund. We analysed and compared two approaches 
In the first analysed model, referred to as FFNS$_{CJKL}$, we adopted a widely 
used massive quark approach in which heavy quark, $h$, contributes to the photon 
structure only through the so-called Bethe-Heitler, 
$\gamma^* \gamma \to h\bar h$ process. In such models the heavy-quark masses are
kept to their physical values. In the second, CJKL model, we used the 
ACOT$(\chi)$ \cite{acot} scheme, where heavy-quark densities appear. It was the 
very first application of this scheme to the photon structure. We performed two 
global fits to the set of updated \fun data collected in various $e^+e^-$ 
experiments. We based both models on the idea of radiatively generated parton 
distributions introduced by the GRV group (see \cite{grv92} for the photonic 
case).

In this work the main assumptions of the previous analysis are left unchanged, 
although some details are improved, and we analyse an additional model. First 
of all, we improve the description of the running of the strong coupling 
constant, $\alpha_s$, and use $\Lambda_{QCD}^{(4)}$ value substantially smaller
than the one used previously. The $\Lambda_{QCD}$ values applied in the former 
analysis were obtained with the assumption that the LO and NLO $\Lambda_{QCD}$ 
values for four active flavours are equal, as in the GRV analysis. They appeared 
to be too high and were one of the causes for the high $\chi^2$ of the fits.

The old FFNS$_{CJKL}$ model is now being denoted as FFNS$_{CJK}$1. We compare 
it with the more realistic model FFNS$_{CJK}$2, in which we include the 
so-called ``resolved-photon'' contribution to \fun given by the 
$\gamma^* G\gam \to h\bar h$ process \cite{grst}. As we stated already in 
\cite{cjkl}, the CJKL type model needs correction which could resolve the 
problem of the high-$x$ behaviour of \fund,  predicted by this model. Therefore,
we analyse a new CJK model in which this problem is avoided. The CJK model is 
further slightly improved by changing the lower limit of the integration, in 
the heavy-quark subtraction terms, from the square of the mass of the heavy 
quark, $m_h^2$, to the starting scale of the DGLAP evolution, $Q_0^2$. This 
also leads to better fits. Finally in the case of the CJK model we abandon the 
valence-number sum rule imposed by hand on the VMD input densities but keep the
corresponding energy-momentum constraint.

Finally, in all new fits we use limited set of data, excluding the TPC2$\gamma$
data which are considered to be inconsistent with other experimental results. 
This is an additional reason, beyond improvements mentioned above, why our new 
predictions for model FFNS$_{CJK}$1 are not identical with results obtained in 
\cite{cjkl} from full set of data.

We also proceed in the direction not addressed before. We analyse uncertainties
of the parton distributions due to the experimental errors of \fun data. This 
part of the work has been motivated by the recent analysis performed for the 
proton structure by the CTEQ Collaboration, \cite{cteq1,cteq2,cteq3} and the 
MRST group, \cite{mrst}. We use the Hessian method, formulated in recent 
papers, to obtain sets of test parton densities allowing along with the 
parton distributions of the best fit to calculate the best estimate and 
uncertainty of any observable depending on the photon structure. Full 
discussion has been given in \cite{uncert}, see also \cite{ph03}.

This paper is divided into four parts. Section 2 recalls the previous
FFNS$_{CJKL}$ and CJKL models of the real photon structure. In section 3 we 
explain in detail all the changes currently introduced to the models. Next, in 
section 4, we present results of the new fits to the experimental \fun data 
along with the calculated uncertainties of the CJK parton distributions. The 
parton distributions which are a result of our analysis have been parametrized 
on the grid. In section 5 we give a short summary and information where to find the corresponding FORTAN programs.


\section{FFNS$_{CJKL}$ and CJKL models - short recollection}

The two approaches leading to the FFNS$_{CJKL}$ and CJKL models, and considered
also in this analysis, have been described in detail in our previous paper 
\cite{cjkl}. The difference between them lies in the approach to the 
calculation of the heavy, charm- and beauty-, quark contributions to the 
photon structure function \fund. First, FFNS$_{CJKL}$ model bases on a widely 
adopted Fixed Flavour-Number Scheme in which there are no heavy quarks (denoted 
below by $h$) as partons in the photon. Their contributions to \fun are given 
by the 'direct' (Bethe-Heitler) $\gamma^* \gamma \to h\bar h$ process. 
In addition one can also include the so-called 'resolved'-photon contribution: 
$\gamma^* G\gam \to h\bar h$. We denote these terms as 
$F_{2,h}\gam(x,Q^2)|_{dir}$ and $F_{2,h}\gam(x,Q^2)|_{res}$, respectively. In 
this paper we consider two FFNS models: in the first one, FFNS$_{CJK}$1, we 
neglect the resolved-photon contribution, while in the second one, 
FFNS$_{CJK}$2, both mentioned contributions to \fun are included. The photon 
structure function is then computed as
\be
\label{F2FFNS}
F_2\gam(x,Q^2) = \sum_{i=1}^3 xe_i^2 (q_i\gam +\bar q_i\gam)(x,Q^2) +
\sum_{h(=c,b)} \left[ F_{2,h}\gam(x,Q^2)|_{dir} + 
F_{2,h}\gam(x,Q^2)|_{res} \right],
\end{equation}
with $q\gam_i(x,Q^2)$ ($\bar q\gam_i(x,Q^2)$) being the light $u,d,s$ quark 
(anti-quark) densities, governed by the LO Dokshitzer-Gribov-Lipatov-Altarelli-Parisi (DGLAP) evolution equations, \cite{DGLAP}.

The CJKL model adopts the new ACOT($\chi$) scheme, \cite{acot}, which is a 
recent realization of the General Variable-Flavour Number Scheme (GVFNS). In 
this scheme one combines the Zero-mass Variable-Flavour Number Scheme (ZVFNS), 
where the heavy quarks are considered as massless partons of the photon, with
the FFNS just discussed above. In this model, in addition to the terms shown 
in Eq. (\ref{F2FFNS}),  one must include the contributions due to the 
heavy-quark densities which now appear also in the DGLAP evolution equations.
A double counting of the heavy-quark contributions to \fun must be corrected 
with the introduction of subtraction terms for both, the direct- and 
resolved-photon, contributions. Further, following the ACOT($\chi$) scheme, we
introduce the $\chi_h=x(1+4m_h^2/Q^2)$ variables giving the proper 
vanishing of the heavy-quark densities at the kinematic thresholds for their 
production in DIS: $W^2=Q^2(1-x)/x>4m_h^2$, where $W$ is the $\gamma^* \gamma$ 
centre of mass energy. Adequate substitution of $x$ with $\chi_h$ in $q_h$ and
the subtraction terms forces their correct threshold behaviour, as 
$\chi_h \to 1$ when $W \to 2m_h$. This is achieved for all terms except for the
direct subtraction term $F_{2,h}\gam(x,Q^2)|_{dir,sub}$ for which
there is a need of an additional condition, 
$F_{2,h}\gam(x,Q^2)|_{dir,sub}=0$ for $\chi_h > 1$. The full formula for 
the photon structure function in the CJKL model is
\ba
F_2\gam(x,Q^2) &=& x\sum_{i=1}^3 e_i^2 (q_i\gam +\bar q_i\gam)(x,Q^2) +
x\sum_{h(=c,b)} e_h^2 (q_h\gam + \bar q_h\gam)(\chi_h,Q^2) \nonumber 
\label{finalacot} \\
&& + \sum_{h(=c,b)} \left[ F_{2,h}\gam(x,Q^2)|_{dir} + 
F_{2,h}\gam(x,Q^2)|_{res} \right]  \\
&& - \sum_{h(=c,b)} \left[ F\gam_{2,h}|_{dir,sub} (\chi_h,Q^2)
      + F\gam_{2,h}|_{res,sub} (\chi_h,Q^2) \right]. \nonumber
\ea
with positivity constraint for each heavy-quark contribution , 
$F_{2,h}\gam(x,Q^2)>0$. Explicit expressions for the terms appearing in 
Eqs. (\ref{F2FFNS}) and (\ref{finalacot}) can be found in \cite{cjkl}.

We use the DGLAP equations summing the QCD corrections in form of leading 
logarithms of $Q^2$. Their solution for quark densities can be divided into 
the so-called point-like (pl) part, equal to a special solution of the full 
inhomogeneous equation and the hadron-like (had) part, arising as a general 
solution of the homogeneous equation. Their sum gives the partonic density in 
the photon. It is most useful to write the result in the Mellin-moments space 
with the $n$th moment defined as 
\be
f^n (Q^2) = \int_0^1 x^{n-1} f(x,Q^2) dx,
\end{equation}
where $f(x,Q^2)$ can be the parton (quark and gluon) densities, 
$q\gam (x,Q^2)$, or the splitting functions in the DGLAP evolution equations 
$P(x,Q^2)$ and $k(x,Q^2)$ defined in \cite{cjkl}. Then one obtains
\be
q^{\gamma,n}(Q^2) = q_{\mathrm{had}}^{\gamma,n}(Q^2) 
+ q_{\mathrm{pl}}^{\gamma,n}(Q^2),
\end{equation}
where
\ba
q_{\mathrm{pl}}^{\gamma,n} (Q^2) &=& \frac{4\pi}{\alpha_s(Q^2)}
\frac{1}{1-2P^{n}/\beta_0}\frac{\alpha}{2\pi \beta_0}
\left[1-L^{1-2P^{n}/\beta_0}\right] k^{n}, \nonumber \\
\quad & \quad & \quad \\
q_{\mathrm{had}}^{\gamma,n} (Q^2) &=& L^{-2P^{n}/\beta_0}q^{\gamma,n}(Q_0^2). 
\label{plhad}
\nonumber
\ea
Here $L= \frac{\alpha_s(Q^2)}{\alpha_s(Q_0^2)}$, where $Q_0^2$ is the scale at 
which the evolution starts (we call it the input scale).

For all models we choose to start the DGLAP evolution at small value of the 
$Q^2$ scale, $Q_0^2=0.25$ GeV$^2$, {following GRV \cite{grv92}} hence our parton
densities are radiatively generated. As it is seen above the point-like 
contributions are calculable without further assumptions, while the hadronic 
parts need input distributions. For this purpose we utilize the Vector Meson 
Dominance (VMD) model \cite{VMD}, where
\be
f_{had}\gam(x,Q_0^2) = \sum_{V}\frac{4\pi \alpha}{\hat f^2_{V}}f^{V}(x,Q_0^2), 
\end{equation}
with the sum running over all light vector mesons (V) into which the photon 
can fluctuate. The parameters $\hat f^2_{V}$ can be extracted from the 
experimental data on the $\Gamma(V\to e^+e^-)$ width. In this analysis we use 
explicitely the $\rho^0$-meson densities while the contributions from other 
mesons are accounted for via a parameter $\kappa$, that is left as a free 
parameter. Thus, we take the parton densities in the photon equal to
\be
f_{had}\gam(x,Q_0^2) = \kappa\frac{4\pi \alpha}{\hat f^2_{\rho}}f^{\rho}(x,Q_0^2).
\label{vmdfor}
\end{equation}

We take the input densities of the $\rho^0$ meson at $Q_0^2=0.25$ GeV$^2$ in
the form of valence-like distributions both for the (light) quark 
($v^{\rho}$) and gluon ($G^{\rho}$) densities. All sea-quark distributions 
(denoted as $\zeta^{\rho}$), including $s$-quarks, are neglected at the input 
scale.

The valence-quark and gluon densities satisfy the energy-momentum sum rule for 
$\rho^0$:
\be
\int_0^1 x(2v^{\rho}(x,Q_0^2)+G^{\rho}(x,Q_0^2))dx = 1,
\label{const2}
\end{equation}
and the sum rule related to the number of valence quarks, $n_v$ 
\be
n_v = \int_0^1 2 v^\rho (x,Q_0^2) = 2.
\label{nv}
\end{equation}
Both of them we imposed as constraints on the parameters of the models in our 
previous analysis \cite{cjkl}.

The input quark and gluon densities are taken in the form (with $\alpha > 0$)
\ba
xv^{\rho}(x,Q_0^2) &=& N_v x^{\alpha}(1-x)^{\beta},  \nonumber \\
xG^{\rho}(x,Q_0^2) &=& \tilde N_g xv^{\rho}(x,Q_0^2)= 
N_g x^{\alpha}(1-x)^{\beta}, \label{input1} \\
x \zeta^{\rho}(x,Q_0^2)&=&0, \nonumber
\ea
where $N_g=\tilde N_gN_v$. The imposed constraint given by Eq. (\ref{const2}) 
allows to express the normalization factor $N_g$ as a function of 
$\alpha, \beta$ and $N_v$. Moreover, when constraint (\ref{nv}) is imposed the 
$N_v$ parameter can be further expressed in terms of $\alpha$ and $\beta$. In 
the former case there are three and in the letter case four free parameters left
as a subject to the global fit to the \fun data.


\section{New analysis}

We have performed new fits with a slightly changed data set as compared to the 
previous work. Moreover, in our new analysis we improved the treatment of the 
running of $\alpha_s$, by differentiating the number of active quarks in the 
running of $\alpha_s$ and in {the} evolution equations, and by using lower 
values of $\Lambda_{QCD}$. We first describe new aspects of our analysis which 
are common to all considered models. Aspects relevant only for the CJK model are 
discussed next.


\subsection{Data Set}

New fits were performed using all existing \fun data, 
\cite{CELLO,PLUTO,JADE,TASSO,TOPAZ,AMY,DELPHI,L3,ALEPH,OPAL,HQ2}, apart from the
old TPC2$\gamma$, \cite{TPC}. In our former global analysis \cite{cjkl} we used 
208 \fun experimental points. Now we decided to exclude the TPC2$\gamma$ data 
from the set because it has been pointed out (see for instance \cite{klasen}) 
that these data are not in agreement with other measurements. A detailed study 
of the influence of various $F_2\gam$ data sets on the fits is performed by
one of us in \cite{uncert}. After the exclusion of the TPC2$\gamma$ data we are 
left with 182 \fun experimental points. We include all these data in the 
$\chi^2$ fit without any weights. A list of all experimental points used can be 
found on the web-page \cite{webpage}.

The exclusion of the TPC2$\gamma$ points affects the $\chi^2/_{\rm DOF}$ of the
fits but has very small influence on the shape of the resulting parton 
distributions.


\subsection{$\alpha_s$ running and values of $\Lambda^{(N_q)}$}

The running of the strong coupling constant at lowest order is given by the 
well-known formula:
\be
\alpha_s^{(N_q)}(Q^2) = \frac{4\pi}{\beta_0 \ln Q^2/(\Lambda^{(N_q)})^2} 
\quad 
\mathrm{with} \quad \beta_0 = 11-\frac{2}{3}N_q,
\label{alphasold}
\end{equation}
where $N_q$ is the number of quarks entering in the $\alpha_s$ evaluation.
\footnote{Notice that we distinguish now between the number of active quarks 
in the photon, denoted by $N_f$, and the number of quarks contributing to the 
running of $\alpha_s$, $N_q$.} This number increases by one unit whenever $Q^2$ 
reaches a heavy-quark threshold, i.e. when $Q^2 = m_h^2$, where the condition 
$\alpha_s^{(N_q)}(m_h^2)=\alpha_s^{(N_q+1)}(m_h^2)$ is imposed in order to 
ensure the continuity of the strong coupling constant.

In our previous analysis $N_f$ was identified with the number of active quarks
in the photon: $N_f=3$ and $5$ in the FFNS$_{CJKL}$ and CJKL models, 
respectively. Since we now distinguish between both numbers of quarks we have to 
use slightly more complicated formulea for the evolution of the parton densities,
as now the above equations depend also on $N_q$ through their dependence on 
$\alpha_s(Q^2)$ and $\beta_0$. Because of the implicit introduction of the 
heavy-quark thresholds into the $\alpha_s$ running we must proceed in three 
steps to perform the DGLAP evolution. In the first step, describing the 
evolution from the input scale $Q_0$ to the charm-quark mass $m_c$, the hadronic 
input $q_{\mathrm{had}}\gam(x,Q_0^2)$ is taken from the VMD model. In the second 
step we evolve the parton distributions from $m_c$ to the beauty-quark mass, 
$m_b$ and in the third one we start at $m_b$. In the second and third steps a 
new hadronic input at $Q^2 = m_h^2$ is given by the sum of the already evolved 
hadronic and point-like contributions and the point-like distribution at this 
scale becomes zero again.

In the previous work we assumed (following the GRV group approach \cite{grv92}) 
that the LO and NLO $\Lambda_{QCD}$ values for four active flavours are equal. 
We adopted $\Lambda_{QCD}^{(4)}=280$ MeV, value given in the Particle Data Group 
(PDG) report \cite{prd}. We now abandon this assumption and take 
$\Lambda_{QCD}^{(4)} = 115$ MeV, which is obtained from the world average value 
$\alpha_s(M_Z) = 0.117$, with $M_Z = 91.188$ GeV, using the LO expression for 
$\alpha_s$ evolution, Eq. (\ref{alphasold}). As a consistency check we performed
fits keeping $\Lambda_{QCD}^{(4)}$ as a free parameter and obtained results 
close to $115$ MeV. Since it is not our aim in this paper to extract a value of 
$\alpha_s$ from a fit to $F_2\gam$ data, we prefer to fix 
$\Lambda_{QCD}^{(4)} = 115$ MeV rather than add a new free parameter in our 
fits. Imposing the continuity condition for the strong coupling constant and 
$m_c = 1.3$ GeV and $m_b = 4.3$ GeV, we obtain $\Lambda_{QCD}^{(3)} = 138$ MeV 
and $\Lambda_{QCD}^{(5)} = 84$ MeV.


\subsection{VMD input}

Like in our previous work the input evolution scale has been chosen to be 
small, $Q_0^2=0.25$ GeV$^2$ for both types of models and we apply the same form 
of the VMD model input, given by Eqs. (\ref{input1}) and Eq. (\ref{vmdfor}).

Finally, we try to relax the constraint on number of the valence quarks, 
$n_v$, in the $\rho$ meson. This leads to 4-parameter fits. That, as will be
quantitatively shown in next sections, is possible only in the case of the
CJK model. Therefore in each of the new FFNS$_{CJK}$ models we have 3 free 
parameters.


\subsection{{Modified subtraction terms in the CJK model}}

In \cite{cjkl} we derived the subtraction term for a direct contribution,
$F_{2,h}\gam|_{dir,sub}$, from the integration of a part of the DGLAP 
evolution equations, namely: 
\be
\frac{dq_h\gam (x,Q^2)}{d\ln Q^2} = \frac{\alpha}{2\pi}e_h^2 k(x),
\end{equation}
where $k(x)$ is the lowest order photon-quark splitting function (see Eq. (7)
in \cite{cjkl}). The question here is: What should the limits of the 
integration be? The upper limit is obviously the $Q^2$ scale at which the
subtraction term is calculated. For the lower limit we took previously the 
standard for the Bethe-Heitler process scale: $Q^2_{low} = m_h^2$. However, the 
threshold condition is $W^2 \leq 4 m_h^2$. This means that even for 
$Q^2 < m_h^2$ the heavy-quark contributions do not vanish as long as the 
condition $x < Q^2/(Q^2+ 4 m_h^2)$ is fulfilled. Hence, in this paper we take
$Q^2_{low}= Q^2_0$ and the direct subtraction term is given by:
\be
F_{2,h}\gam|_{dir,sub}(x,Q^2)= x \ln \frac{Q^2}{Q_0^2}  
3e_h^4 \frac{\alpha}{\pi}(x^2+(1-x)^2).
\label{dirsub2}
\end{equation}
The same discussion applies to the subtraction term for the resolved-photon  
contribution. So we now use
\be
F_{2,h}\gam|_{res,sub}= x \ln \frac{Q^2}{Q_0^2}
e_h^2 \frac{\alpha_s(Q^2)}{\pi}\int_{x}^1 \frac{dy}{y}
P_{qG}(\frac{x}{y})G\gam(y,Q^2).
\label{ss19}
\end{equation}
instead of Eq. (19) in \cite{cjkl}. We found that the quality of the fit 
improves with this choice of the $Q^2_{low}$.

As we noticed in Section 2 the $x\to \chi_h$ substitution leads to the proper 
threshold behaviour of all the heavy-quark contributions to the \fund, except for
the subtraction term for direct contribution. It is already seen in 
Eq. (\ref{dirsub2}) that this term does not vanish for $\chi_h \to 1$ and 
therefore by subtracting it the resulting heavy-quark contribution to 
$F_2\gam$ may become negative in some regions of the $x$ and $Q^2$ plane. 
An extra constraint to avoid this unphysical situation is, thus, needed. In 
Ref. \cite{cjkl} we imposed the simple condition (positivity constraint) that 
the heavy-quark contribution to \fun has to be positive. Unfortunately, this 
constraint was not strong enough and for some small windows at small and large 
$x$ still the unphysical situation $F_{2,h}\gam(x,Q^2)<F_{2,h}\gam(x,Q^2)|_{dir}+
F_{2,h}\gam(x,Q^2)|_{res}$ was found \cite{MariuszPrzybycien}. Therefore, in 
this analysis we apply a positivity condition in the following form: 
\be 
F_{2,h}\gam(x,Q^2) \geq F_{2,h}\gam(x,Q^2)|_{dir}
                      + F_{2,h}\gam(x,Q^2)|_{res}.
\end{equation}


\section{Results of the new $F_2\gam$ global fits}

In this analysis we determine the parameters of the models, related to the 
initial quark and gluon densities at the scale $Q^2_0=0.25 \; GeV^2$, by means 
of the global fits to the experimental data on \fund. We use 182 \fun 
experimental points, 
\cite{CELLO,PLUTO,JADE,TASSO,TOPAZ,AMY,DELPHI,L3,ALEPH,OPAL,HQ2}, with 
equal weights. Still, it has been shown in \cite{uncert} that when we remove the
CELLO \cite{CELLO} and DELPHI 2001 \cite{DELPHI} data sets the quality of the 
fit improves substantially but the parton distributions lie well within the CJK 
uncertainties. However, we think that there is no strong argument to discard 
these data sets. So, we have chosen to keep them with the same weight as the 
other sets. Fits are based on the least-squares principle (minimum of $\chi^2$) 
and were done using \textsc{Minuit} \cite{minuit}. Systematic and statistical 
errors on data points were added in quadrature.

In the CJK model we have four free parameters: $\alpha,\beta,N_v,\kappa$, 
Eqs. (\ref{input1}) and Eq. (\ref{vmdfor}). On the other hand, the two FFNS 
models differ only in the inclusion or not of the resolved-photon contribution 
to \fund (only the FFNS$_{CJK}$2 model takes it into account through the 
$\gamma^* G\gam \to h \bar h$ process). For both FFNS models we impose 
the number of valence quarks constraint (\ref{nv}) that allows to express $N_v$ 
in terms of $\alpha$ and $\beta$ reducing the number of free parameters to three.

The parameters of our new fits are presented in Table \ref{tparam}. The second 
and third columns show the quality of the fits, i.e. the total $\chi^2$ for
182 points and the $\chi^2$ per degree of freedom. The fitted values for
parameters $\alpha$, $\beta$, $\kappa$ and $N_v$ are presented in the middle
of the table with the errors obtained from \textsc{Minos} with the standard 
requirement of $\Delta \chi^2 = 1$. In the last column the value obtained from 
the constraint (\ref{const2}) for $\tilde N_g$ from other parameters is given.

In the case of the FFNS$_{CJK}$ models the test fits with the abandoned 
constraint (\ref{nv}) gives $n_v \approx 0.5$ and $\approx 1.4$ in the 
FFNS$_{CJK}$1 and FFNS$_{CJK}$2 models, respectively. We think that it is 
too far away from the expected value $n_v = 2$ \footnote{Note, that in the CJK 
model we obtain $n_v = 2.0\pm 0.1$}. This is the reason for keeping the 
constraint (\ref{nv}) for both FFNS$_{CJK}$ models. In this case the $N_v$ 
parameter is calculated from the constraint (\ref{nv}) and therefore we do not 
state its error.

\begin{table}[htb]
\begin{center}
\renewcommand{\arraystretch}{1.5}
\begin{tabular}{|c|@{} p{0.05cm} @{}|c|c|@{} p{0.08cm} @{}|c|c|c|c|@{} p{0.08cm} @{}|c|}
\hline
 model           && $\chi^2$ (182 pts) & $\chi^2/_{\rm DOF}$ && $\kappa$ & $\alpha$ & $\beta$ & $N_v$ && $ \tilde N_g$ \\
\hline
\hline
FFNS$_{CJK}$1 &&        314.0       &         1.754       &&  $2.267^{+0.063}_{-0.072}$ & $0.265^{+0.038}_{-0.032}$ & $0.792^{+0.189}_{-0.149}$ & 0.358 && 5.02  \\
\hline
FFNS$_{CJK}$2 &&        279.8       &         1.563       &&  $2.110^{+0.084}_{-0.090}$ & $0.310^{+0.054}_{-0.051}$ & $0.823^{+0.265}_{-0.223}$ & 0.415 && 4.51  \\
\hline
 CJK           &&        273.7       &         1.537       &&  $1.934^{+0.131}_{-0.124}$  & $0.299^{+0.077}_{-0.069}$ & $0.898^{+0.316}_{-0.275}$ & $0.404^{+0.116}_{-0.088}$ && 4.93  \\
\hline
\end{tabular}
\caption{Results of the fits for the three models considered in the analysis.
The quoted errors are obtained from \textsc{MINOS} with the standard requirement 
of $\Delta \chi^2 = 1$. 
\label{tparam}}
\end{center}
\end{table}

The $\chi^2$ per degree of freedom presented in Table \ref{tparam} are still 
rather high, however there is an improvement as compared with previous fits. The
old $\chi^2/_{\rm DOF}$ for the same set of 182 data points read 1.99 in the 
FFNS$_{CJKL}$ and 1.80 in the CJKL model. We see that the corresponding new 
$\chi^2/_{\rm DOF}$ values are about 0.25 lower. This is mostly due to the 
adoption of much lower $\Lambda_{QCD}$ values as well as the modification of the
subtraction contributions in the CJK model. The rejection of the TPC2$\gamma$ 
data also reduces the value of $\chi^2/_{\rm DOF}$.

We observe that the $\chi^2/_{\rm DOF}$ for the FFNS$_{CJK}$2 and CJK models 
are very similar and lower than the one for the FFNS$_{CJK}$1 model. It is 
obvious that the inclusion of the resolved $\gamma^* G\gam \to h\bar h$ 
contribution to \fund improves the agreement between the model and the data (see
for instance \cite{grs}).

We see that the quality of the present data does not allow for a clear 
discrimination between the different ways of dealing with the heavy quarks 
as the $\chi^2$ and all fitted parameters are very similar. The $\kappa$ values 
are close to 2 which is in agreement with the GRV LO \cite{grv92} prediction. 
The $\alpha$ parameter varies from about 0.25 to 0.31 
($\alpha-1 \approx -0.75,-0.7$) which seems to be conformable with the Regge 
theory predicting that for a valence-quark density $\alpha-1 \approx -0.5$. The 
$\beta$ values ranging from about 0.8 to 0.9 are again in good agreement with 
the GRV LO \cite{grvp} finding (0.85). They are not as small as in the case of 
our former FFNS$_{CJKL}$ model but are far from 2, a standard prediction from 
the quark-counting rule \cite{joffe}.


\subsection{Comparison of the CJK and FFNS$_{CJK}1,2$ fits with the $F_2\gam$ 
data}

In this paper we are going to present plots only for the three models that we 
analyse, without any comparison with other parametrizations. However, we will 
present plots of tthe same type as in Ref. \cite{cjkl} in order to facilitate 
the comparison. Moreover we will describe differences between of our new results
and the previous ones and with the GRS LO \cite{grs} and SaS1D \cite{sas} 
parametrizations. If experimental points for a few values of $Q^2$ are displayed
in a panel, the average of the smallest and biggest one was taken in the 
computation of the theoretical prediction.

Figures \ref{fit1}--\ref{fit4} show a comparison of the CJK and FFNS$_{CJK}$ 
fits to the \fun with the experimental data as a function of $x$, for different 
values of $Q^2$. The FFNS$_{CJK}$1 fit predictions are very similar to the GRS 
LO parametrization results in the whole range of $x$ while the FFNS$_{CJK}$2 and
the CJK model predict a much steeper behaviour of the \fun at small $x$ with 
respect to the FFNS$_{CJK}$1 fit (and GRS LO) and SaS1D parametrizations. On the 
other hand these curves are less steep than the old FFNS$_{CJKL}$ and CJKL ones.
The behaviour of the three fits in the region $x \gtrsim 0.1$, as shown in 
Figs. \ref{fit3} and \ref{fit4}, is very similar.

Apart from this direct comparison with the photon structure function
$F_2\gam$ data, we perform another comparison, this time with LEP data 
that were not used directly in our analysis. Figures \ref{evol1} and 
\ref{evol2} present our predictions for the \fund, averaged over various $x$
regions, compared with the recent OPAL data \cite{HQ2}. Like in our previous 
analysis we see that all FFNS type predictions (including GRS LO and SaS1D 
parametrizations) are similar and fairly well describe the experimental data. 
Moreover, again the CJK model, alike the CJKL model, slightly differs from 
other fits. However, this difference is much smaller now and gives better 
agreement with the data.

We observe that for the case of the medium-$x$ range, $0.1<x<0.6$, there are 
small differences between the CJK and both FFNS models at very low $Q^2$, where 
there are no experimental data, and at large $Q^2$ where a slightly better 
agreement between the CJK model prediction and the OPAL data is found. Comparing
with the plot in Fig. \ref{evol2} we see that larger values of the \fun obtained 
for the CJK model are originated at lower values of $x$ in the considered range,
as one could expect from  Figs. \ref{fit1}--\ref{fit4}.


\subsection{Parton densities}

It is very instructive to discuss the parton densities obtained in the CJK 
and FFNS$_{CJK}$ fits. We choose to present results for medium- and high-$x$ at 
$Q^2=10$ GeV$^2$, see  Fig. \ref{parton}. In Figs. \ref{updens}--\ref{chmdens} 
we show the up-, and charm-quark and gluon densities for various $Q^2$ values. 
First we observe that parton distributions obtained in various models are very 
similar, except that of course there are no heavy-quark distributions in 
FFNS-type approaches. In the case of the CJK models, due to the introduction of 
the $\chi_h$ variable, the $c\gam(x,Q^2)$ and $b\gam(x,Q^2)$ densities vanish 
not at $x=1$, as in the case of the GRV LO \cite{grv92} and SaS1D 
parametrizations, but as they should at the kinematic threshold. Moreover, the 
CJK heavy-quark densities at low $x$ are larger than the corresponding densities
obtained in other parametrizations. This is a feature that can be observed in a 
wide range of $Q^2$ values in Fig. \ref{gludens}.

We notice that our new parton densities have all similar shapes to the 
corresponding old CJKL distributions. Though, in the medium- and high-$x$ 
regions they have slightly higher values. On contrary, at very low $x$ values
CJKL densities are greater than the new ones, opposite case occurs in the 
$10^{-3}<x<10^{-2}$ region. In the case of the gluon density we find that all 
new lines are much steeper at high-$x$ than the predictions of our previous 
models and the GRV LO and SaS1D parametrizations.


\subsection{Comparison with $F\gam_{2,c}$}

We compare the individual contributions included in the CJK model relevant for 
our predictions of the $F_{2,c}\gam$. Results from the CJK fit are presented 
in Fig. \ref{acot} for $Q^2=5,20,100$ and 1000 GeV$^2$. Almost all the
contributing terms vanish in the $W\to 2m_c$ threshold in a natural way
due to the introduction of the $\chi_h$ variable. The only exception is the 
one subtraction term, namely $F_{2,c}|_{dir,sub}$ which dominates near the 
highest kinematically allowed $x$ value and vanishes only due to the extra 
condition, $F_{2,c}|_{dir,sub}=0$ for $\chi_h > 1$, we imposed. The direct 
(Bethe-Heitler) term is important in the medium-$x$ range. Its shape resembles 
the valence-type distribution. The charm-quark density contribution, \ie the 
term $2xe_c^2 c\gam(x,Q^2)$, is important in the whole kinematically 
available $x$ range and dominates the $F_{2,c}\gam$ for small $x$. In this 
region also both resolved-photon contributions increase with decreasing $x$,
but to great extent they cancel each other.

Finally we see that imposing the improved positivity condition
$F_{2,h}\gam>F_{2,h}(x,Q^2)\gam(x,Q^2)|_{dir}+F_{2,h}\gam(x,Q^2)|_{res}$ 
on the heavy-quark contributions to the \fun results in correct threshold 
behaviour of the total $F_{2,c}$ function. Unlike in the case of the CJKL fit, 
the $F_{2,h}\gam$ and its contributions vanish at the same high $x$ value.

\bigskip

A good test of the charm-quark contributions is provided by the OPAL 
measurement of the $F_{2,c}\gam$, obtained from the inclusive production of 
$D^{*\pm}$ mesons in photon-photon collisions \cite{F2c}. The averaged 
$F_{2,c}\gam$ has been determined in the two $x$ bins. These data points are 
compared to the predictions of the CJK and FFNS$_{CJK}$ models and GRS LO and 
SaS1D parametrizations in Fig. \ref{fF2c}.

Our first observation is the following: our models containing the 
resolved-photon contribution, $F_{2,h}\gam(x,Q^2)|_{res}$ 
(FFNS$_{CJK}$2 and the CJK model) agree better with the low-$x$ experimental 
point than other predictions. The GRS LO and SaS1D parametrizations also 
include the resolved-photon term but in their case the gluon density increased
less steep than our models predict, as was already mentioned. Their 
$F_{2,c}\gam$ lines lie below the results of our new fits but higher than the 
FFNS$_{CJK}$1 curve, given solely by the direct Bethe-Heitler contribution.

The CJK model overshoots the experimental point at high $x$ while other 
predictions agree with it within its uncertainty bounds. Taking into account 
both data points we conclude that the best agreement with the experimental 
results is provided by the FFNS$_{CJK}$2 model.


\subsection{Gluon densities at HERA}

We also checked that the gluon densities of CJK and FFNS$_{CJK}$ models agree 
with the H1 measurement of the $G\gam$ distribution performed at 
$Q^2=74$ GeV$^2$ \cite{h1glu}. As can be seen in Fig. \ref{gluh1} all models
predict gluon densities that lie above the one provided by the GRV LO 
parametrization, which gave so far best agreement with the H1 data.
Further comparison of our gluon densities to the H1 data cannot be performed 
in a fully consistent way, since the GRV LO proton and photon parametrization 
were used in the experiment in order to extract such gluon density.


\subsection{The uncertainties of the CJK parton distributions}

Let us now present the main results of an analysis of the CJK parton 
distribution uncertainties described in detail in \cite{uncert}.

During the last two years numerous analysis of the uncertainties of the proton
parton densities resulting from the experimental data errors appeared. The 
CTEQ Collaboration in a series of publications, \cite{cteq1,cteq2,cteq3}, 
developed and applied a new method of their treatment significantly improving 
the traditional approach to this matter. Later the same approach has been 
applied by the MRST group in \cite{mrst}. The method, denoted as Hessian 
method, bases on the Hessian formalism. We applied it for the very first time 
to the case of the photon parton distributions \cite{uncert,ph03}.

The uncertainties analysis was performed for the CJK photon parametrization 
only. The set of the best values of parameters $\kappa, \alpha, \beta$ and 
$N_v$, corresponding to the minimal $\chi^2$ of the global fit, $\chi_0^2$
(table \ref{tparam}), is denoted as the $S^0$ parametrization. Using the Hessian
method we created an additional basis of the test parametrizations of the CJK 
parton densities, $\{S_k^{\pm},k=1,\cdots,4\}$, where 4 corresponds to the 
number of free parameters of the model. The set $\{S_k^{\pm}\}$ allows for the 
calculation of the uncertainty of any physical observable $X$ depending on the 
parton densities. Its best value is given as $X(S^0)$. The uncertainty of $X$,
for a displacement from the parton densities minimum by $\Delta \chi^2 = T^2$ 
(T - the tolerance parameter) can be calculated with a very simple expression 
(named as master equation by the CTEQ Collaboration)
\be
\Delta X = \frac{T}{2t}\left( \sum_{k=1}^d[X(S_k^+)-X(S_k^-)]^2 \right)
^{\frac{1}{2}},
\label{master}
\end{equation}
where parameter $t=5$ in that particular case. After a detailed test of the
allowed deviation of the global fit from the minimum, we found that $T$ should 
lie in the range $5\sim 10$ \cite{uncert}. Note that having calculated 
$\Delta X$ for one value of the tolerance parameter $T$ we can obtain the 
uncertainty of $X$ for any other $T$ by simple scaling of $\Delta X$. This way 
sets of $\{S_k^{\pm}\}$ parton densities give us a perfect tool for studying of 
the uncertainties of other physical quantities. One of such quantities can be 
the parton densities themselves.

In Figure \ref{densuncup} the up-, strange- and charm-quark and gluon densities
calculated in the FFNS$_{CJK}$ models and the GRV LO, GRS LO and SaS1D 
parametrizations are compared with the CJK predictions. We plot for $Q^2=10$ 
and 100 GeV$^2$ the ratios $q\gam(\mathrm{Other \: model})/q\gam(\mathrm{CJK})$ 
and $q\gam(\mathrm{Other \: parametrization})/q\gam(\mathrm{CJK})$ of the parton
$q\gam$ densities calculated in the CJK model and other models (or 
parametrizations). Solid lines show the CJK fit uncertainties for 
$\Delta \chi^2 = 25$ computed with the $\{S_k^{\pm}\}$ test parametrizations. 
First we notice that predictions of FFNS$_{CJK}$ models in the case of all 
parton distributions lie between the lines of the CJK uncertainties. There is 
only one range of $x$, namely $0.01\simkl x \simkl 0.1$ at $Q^2=10$ GeV$^2$ 
where the up- and down-quark densities predicted by the FFNS$_{CJK}$ 1 fit go 
slightly beyond the uncertainty band. This indicates that the choice of 
$\Delta \chi^2 = 25$ agrees with the differences among our four models. 
Moreover, the GRV LO parametrization predictions are nearly contained within the
CJK model uncertainties. Obviously, that is not the case, for the heavy-quark 
densities. The SaS1D results differ very substantially from the CJK ones. Next, 
we observe that, as expected, the up-quark distribution is the one best 
constrained by the experimental data, while the greatest uncertainties are 
connected with the gluon densities. In the case of $u\gam$, the 
$\Delta \chi^2 = 25$ band widens in the small $x$ region. Alike in the case of 
other quark uncertainties it shrinks at high $x$, this is due to the large 
$u$-quark density in the photon at high $x$. On contrary the gluon distributions 
are least constrained at the region of $x\to 1$. Finally we observe that all 
uncertainties become slightly smaller when we go to higher $Q^2$ from 10 to 100 
GeV$^2$ (not shown).


\section{Summary}

We enlarged and improved our previous analysis \cite{cjkl}. Here we performed
3 new global fits to the \fun data, excluding the TPC2$\gamma$ experiment. Two 
additional models were analysed. New fits gave $\chi^2$ per degree of freedom,
1.5-1.7, about 0.25 better than the old results. All features of the CJKL model,
such as the shape of the heavy-quark distributions, good description of the LEP 
data on the $Q^2$ dependence of the $F_2\gam$ and on $F_{2,c}\gam$ are preserved.
We checked that the gluon densities of our models agree with the H1 measurement 
of the $G\gam$ distribution performed at $Q^2=74$ GeV$^2$ \cite{h1glu}.

An analysis of the uncertainties of the CJK parton distributions due to the 
experimental errors based on the Hessian method was performed for the very 
first time for the photon \cite{uncert,ph03}. We constructed set of test 
parametrizations for the CJK model. It allows to computate uncertainties of any 
physical quantity depending on the real photon parton densities.

Fortran parametrization programs for all models, obtained through 
parametrization of the fit results on the grid, can be obtained from the 
web-page \cite{webpage}. Set of the data used in the fits is also given there.


\section*{Acknowledgments}

P.J. would like to thank R.Nisius for important comments, J.Jankowska and 
M.Jankowski for their remarks on the numerical method applied in the grid 
parametrization program and A.Zembrzuski for further useful suggestions. 
We all thank Mariusz Przybycie\'n for his important remark.

This work was partly supported by the European Community's Human Potential 
Programme under contract HPRN-CT-2000-00149 Physics at Collider and 
HPRN-CT-2002-00311 EURIDICE. FC also acknowledges partial financial
support by MCYT under contract FPA2000-1558 and Junta de Andaluc{\'\i}a group 
FQM 330. This work was partially supported  by the Polish Committee for 
Scientific Research, grant  no.~1~P03B~040~26 and project 
no.~115/E-343/SPB/DESY/P-03/DWM517/2003-2005.



\include{figura}

\end{document}

%% file: figura.tex
\clearpage

\begin{figure}
\includegraphics[scale=1.0]{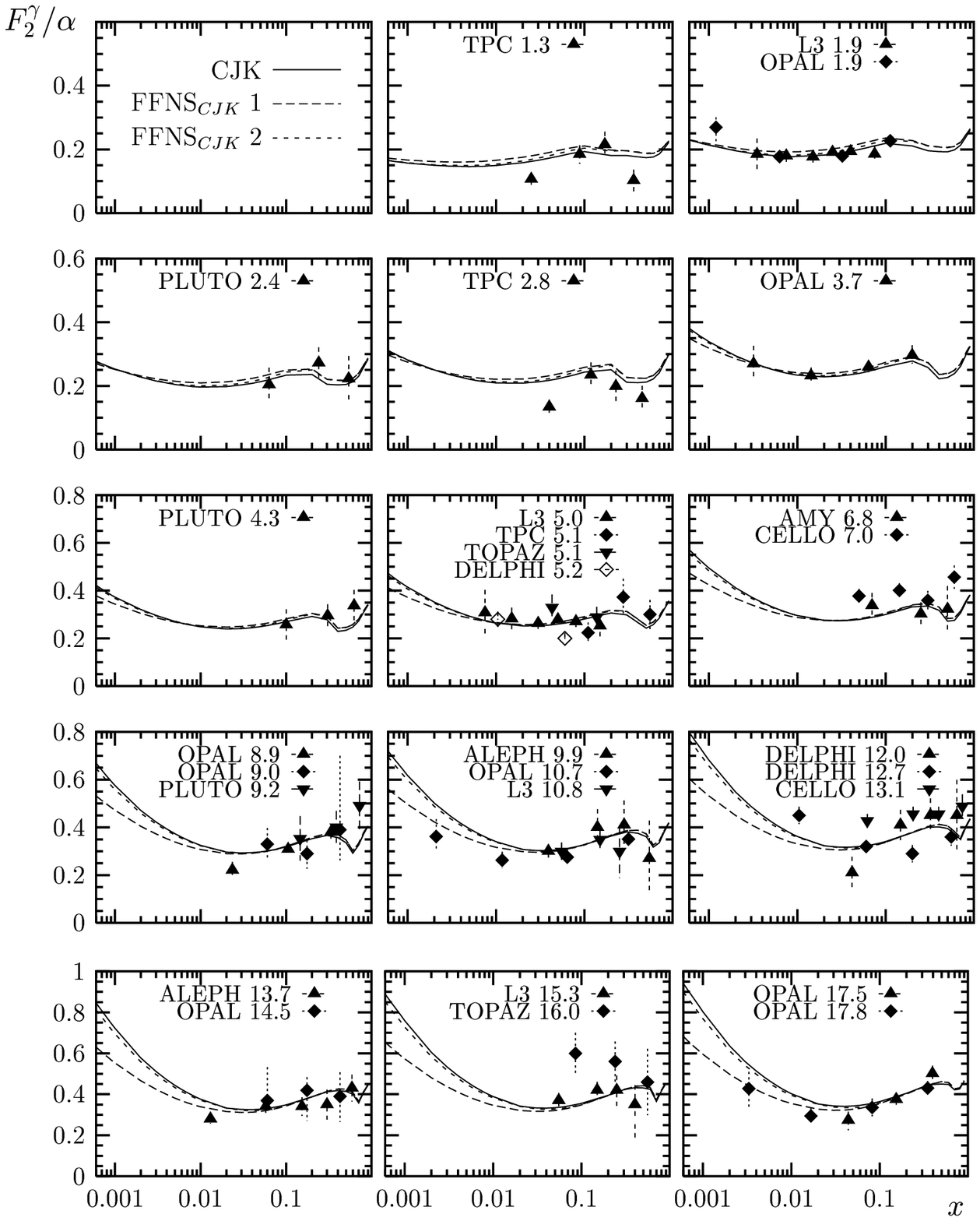}%
\vskip -0.7cm
\caption{Predictions for the $F_2^{\gamma}(x,Q^2)/\alpha$ for the CJK 
and FFNS$_{CJK}$ models compared with the experimental data 
\cite{CELLO,PLUTO,JADE,TASSO,TOPAZ,AMY,DELPHI,L3,ALEPH,OPAL,HQ2}, for small and 
medium $Q^2$ as a function of $x$ (logarithmic scale). If a few values of $Q^2$ 
are displayed in the panel, the average of the smallest and biggest $Q^2$ was 
taken in the computation.}
\label{fit1}
\end{figure}

\newpage

\begin{figure}
\begin{center}
\includegraphics[scale=1.0]{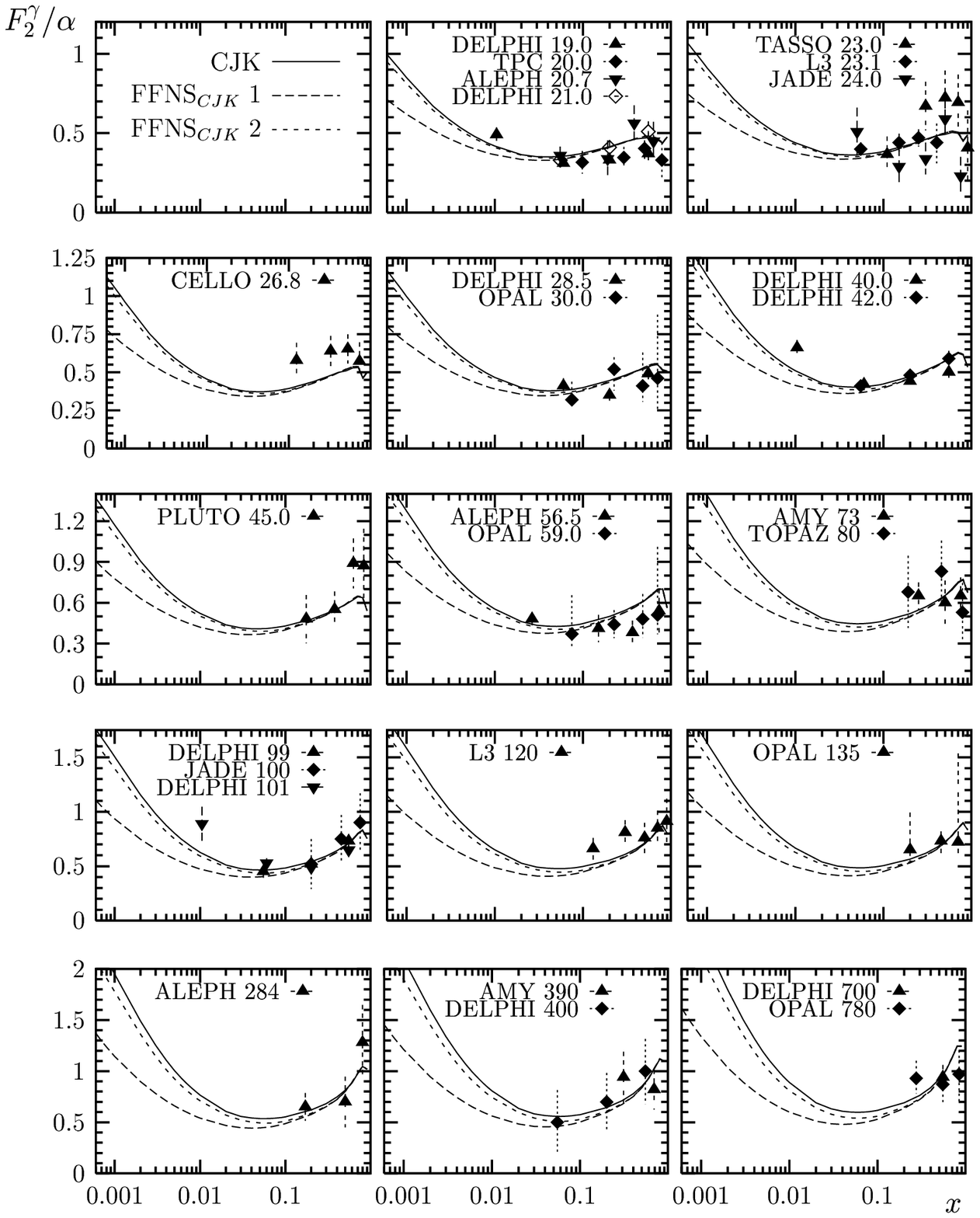}%
\caption{The same as in Fig. \ref{fit1}, for $Q^2 \gtrsim 20 \mathrm{GeV}^2$.}
\label{fit2}
\end{center}
\end{figure}

\newpage

\begin{figure}
\includegraphics[scale=1.0]{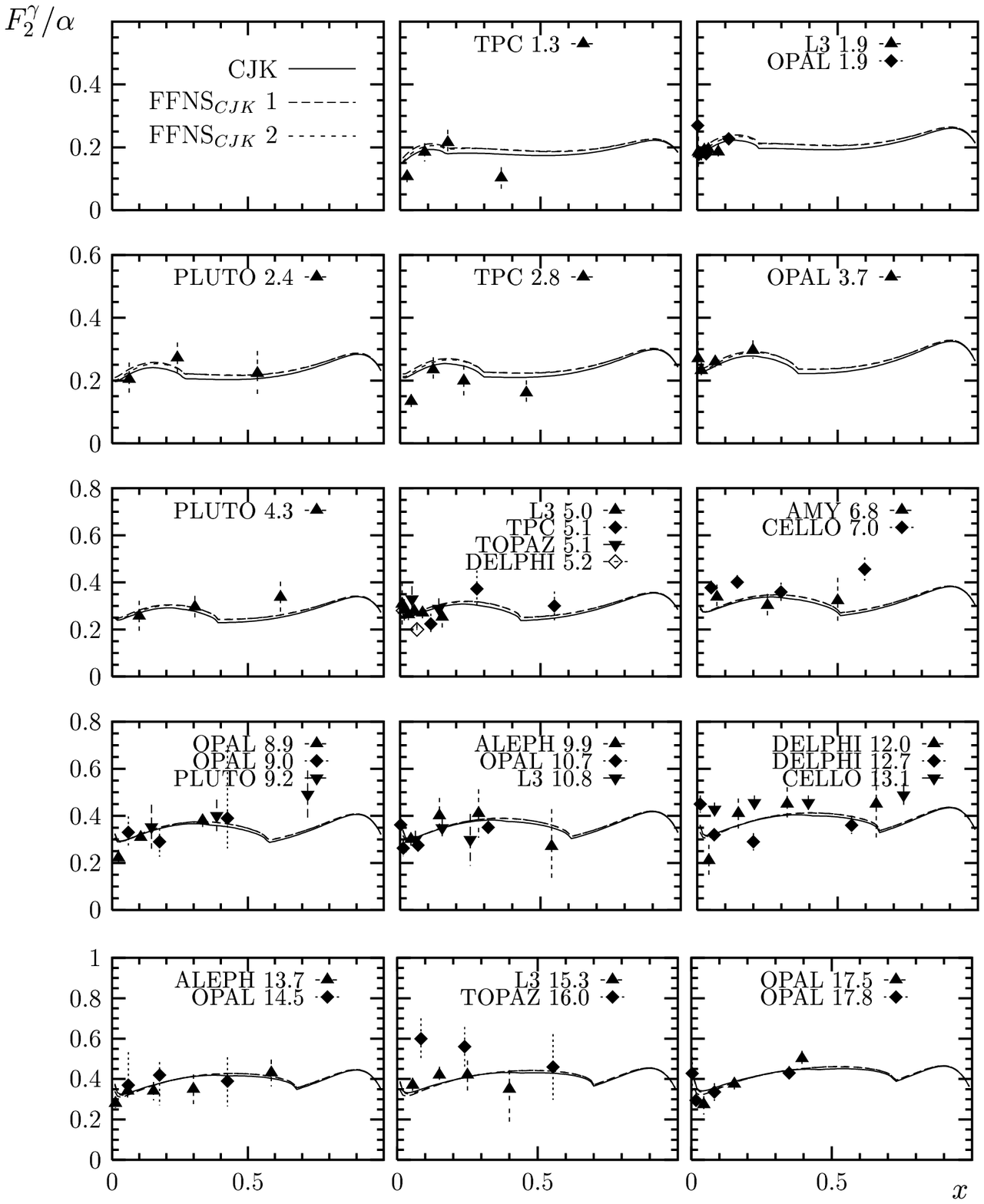}%
\caption{The same as in Fig. \ref{fit1} for a linear scale in $x$.}
\label{fit3}
\end{figure}

\newpage

\begin{figure}
\includegraphics[scale=1.0]{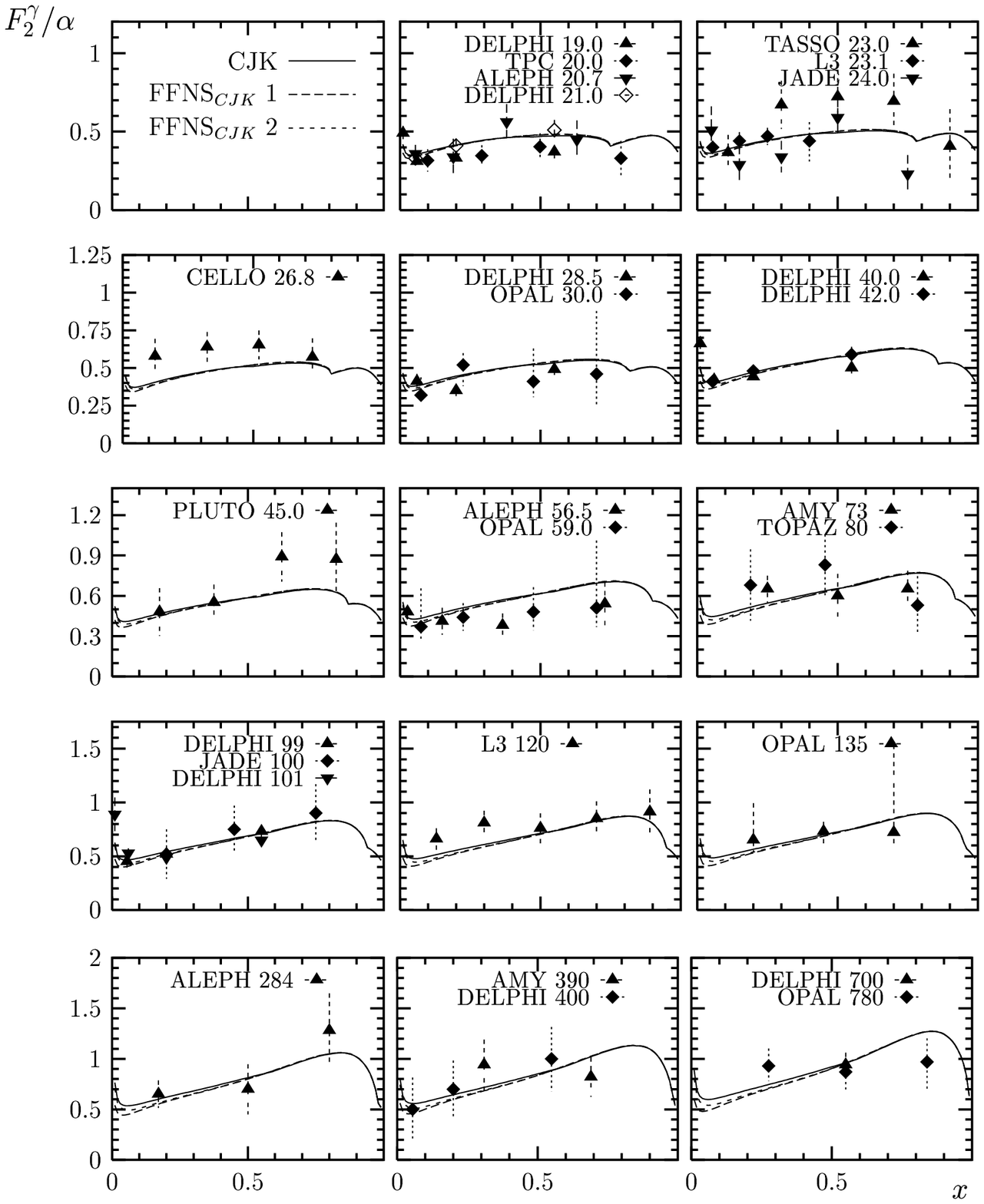}%
\caption{The same as in Fig. \ref{fit2} for a linear scale in $x$.}
\label{fit4}
\end{figure}

\newpage

\begin{figure}
\includegraphics[scale=1.0]{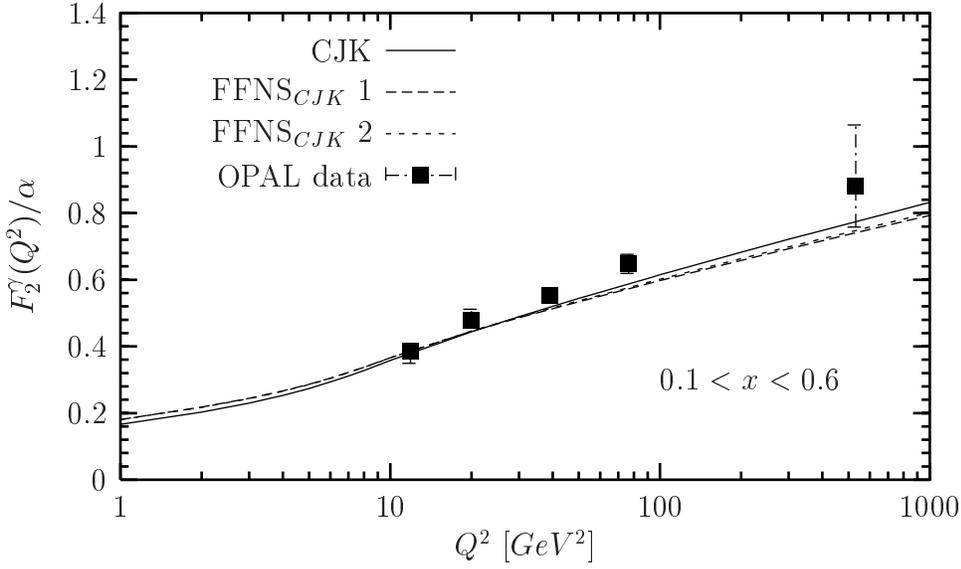}
\caption{Comparison of the recent OPAL data \cite{HQ2} for the
$Q^2$-dependence of the averaged over $0.1<x<0.6$ $F_2^{\gamma}/\alpha $
with the predictions of the CJK and FFNS$_{CJK}$ models.}
\label{evol1}
\end{figure}

\begin{figure}[h]
\includegraphics[scale=1.0]{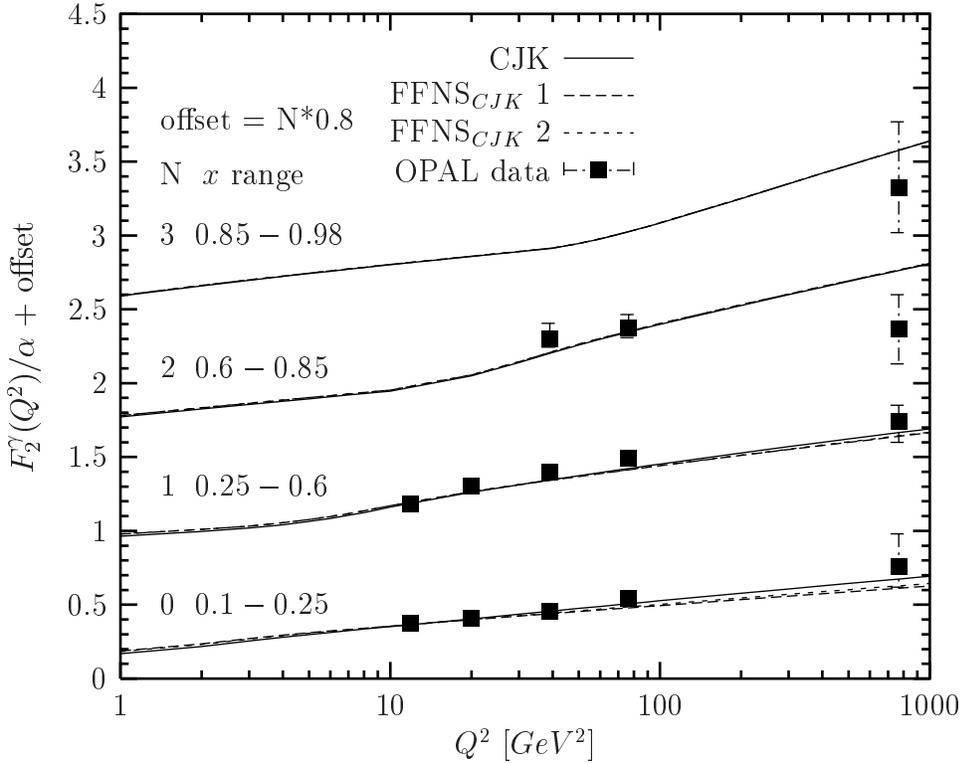}
\caption{
As in Fig.~\ref{evol1} for \fund/$\alpha$, averaged over four different $x$ 
ranges.}
\label{evol2}
\end{figure}

\newpage

\begin{figure}
\begin{center}
\includegraphics[scale=1.0]{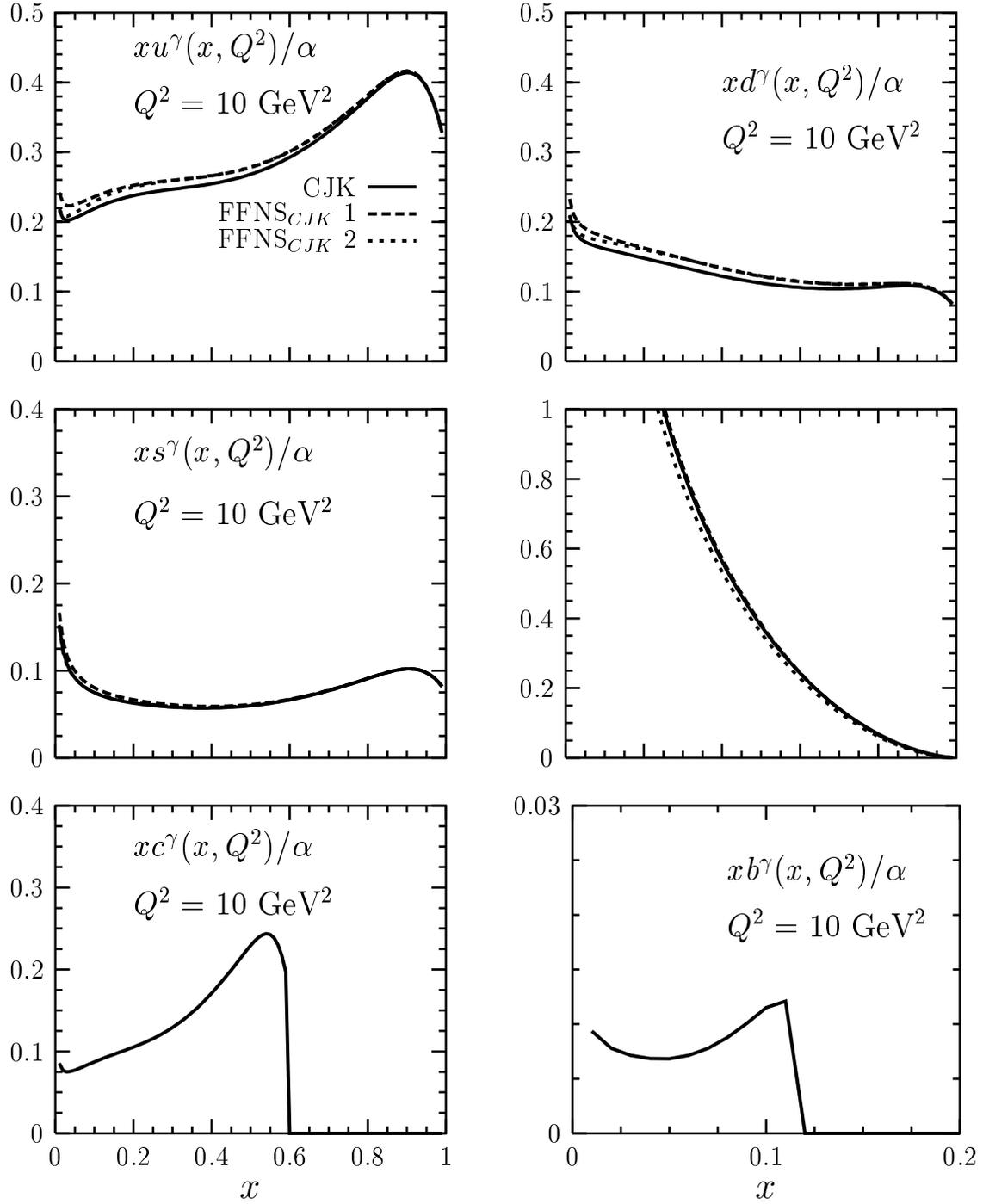}%
\caption{Comparison of the parton densities predicted by various 
models at $Q^2=10$ GeV$^2$, as a function of $x$.}
\label{parton}
\end{center}
\end{figure}

\newpage

\begin{figure}
\begin{center}
\includegraphics[scale=1.0]{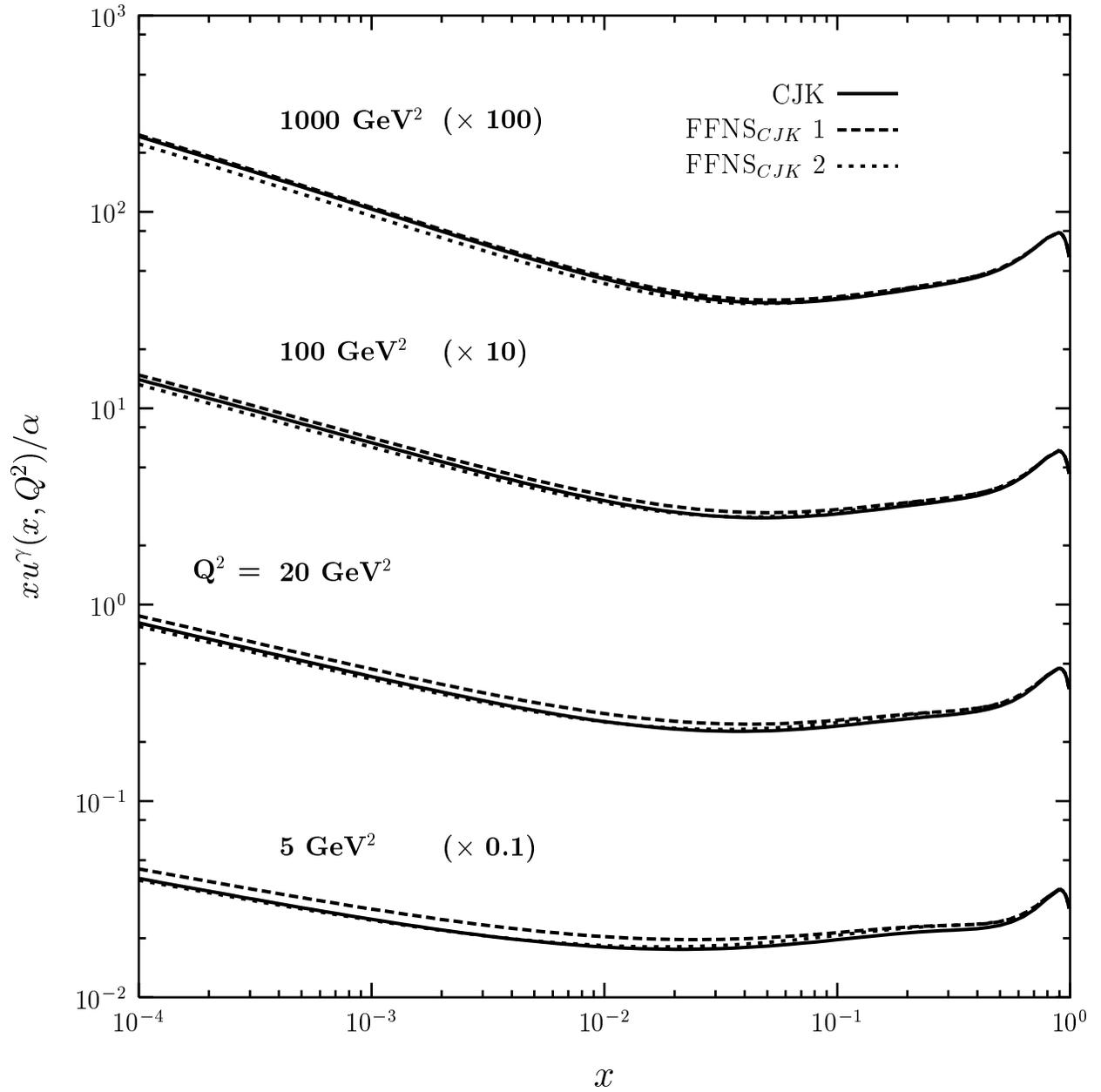}%
\caption{Comparison of the up-quark density at four values of $Q^2$ in the CJK
and FFNS$_{CJK}$ models, as a function of $x$.}
\label{updens}
\end{center}
\end{figure}

\newpage

\begin{figure}
\begin{center}
\includegraphics[scale=1.0]{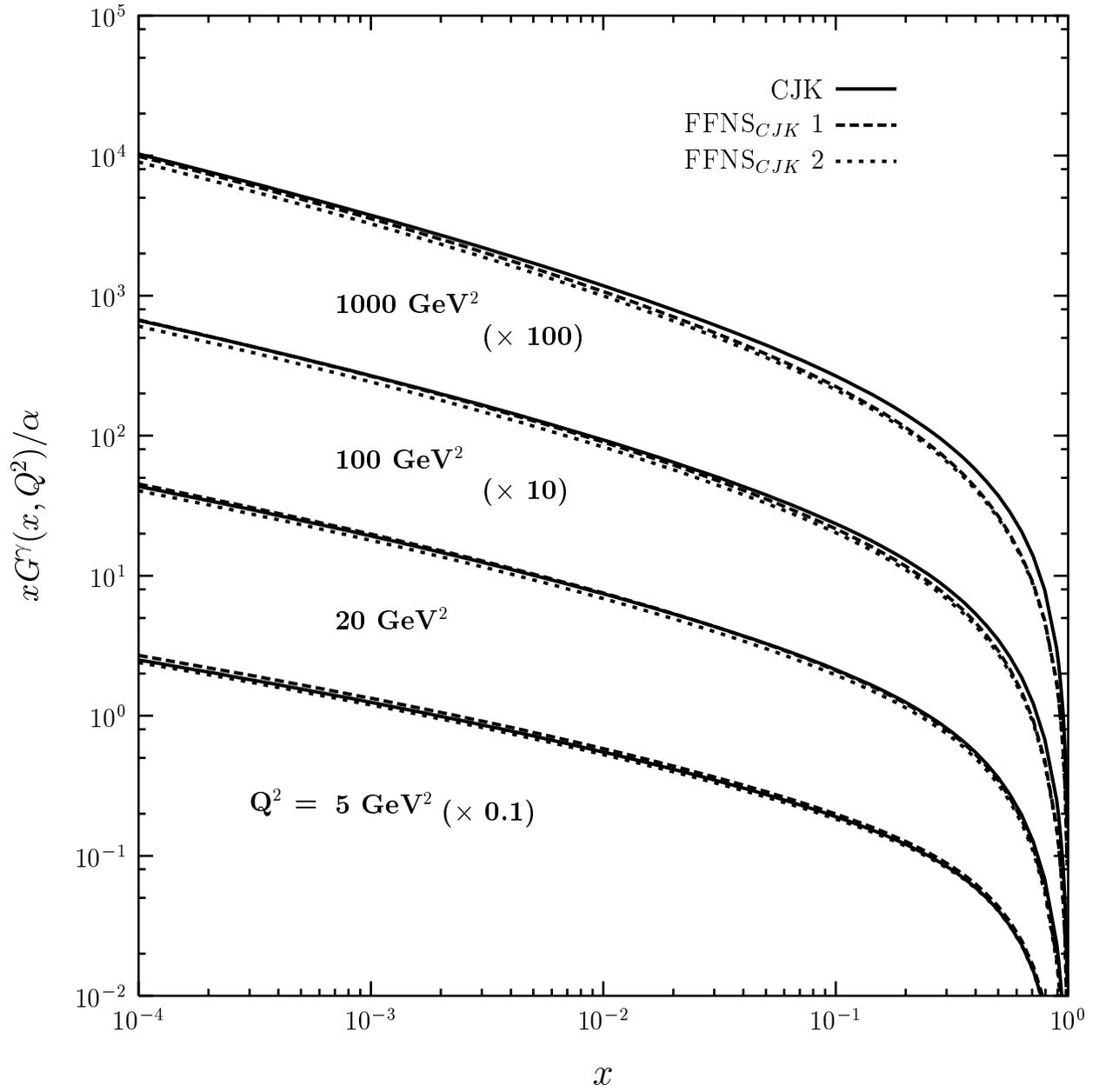}%
\caption{The same as in Fig. \ref{updens} for the gluon density.}
\label{gludens}
\end{center}
\end{figure}

\newpage

\begin{figure}
\begin{center}
\includegraphics[scale=1.0]{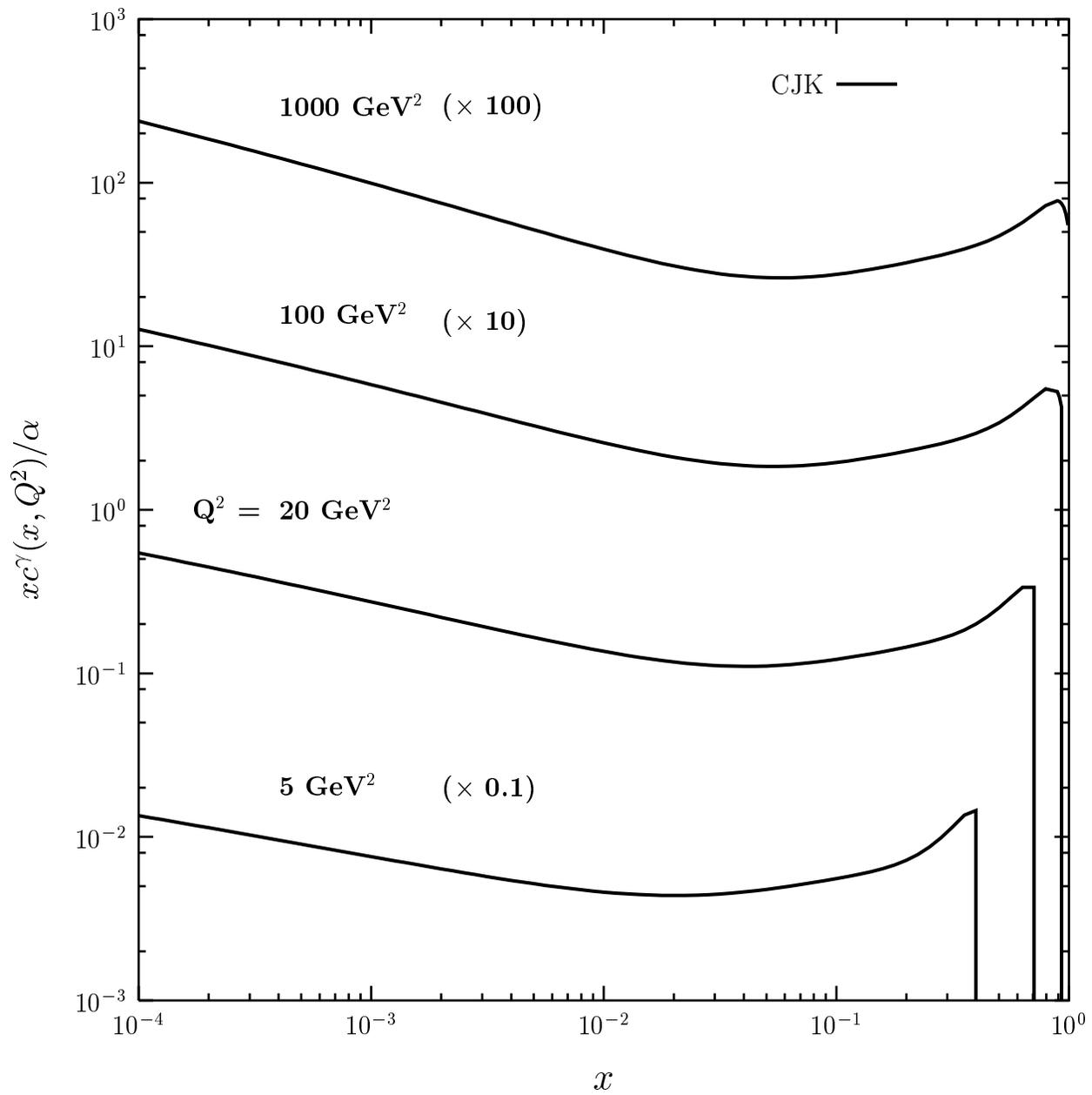}%
\caption{The same as in Fig. \ref{updens} for the charm-quark density.}
\label{chmdens}
\end{center}
\end{figure}

\newpage

\begin{figure}
\includegraphics[scale=1.0]{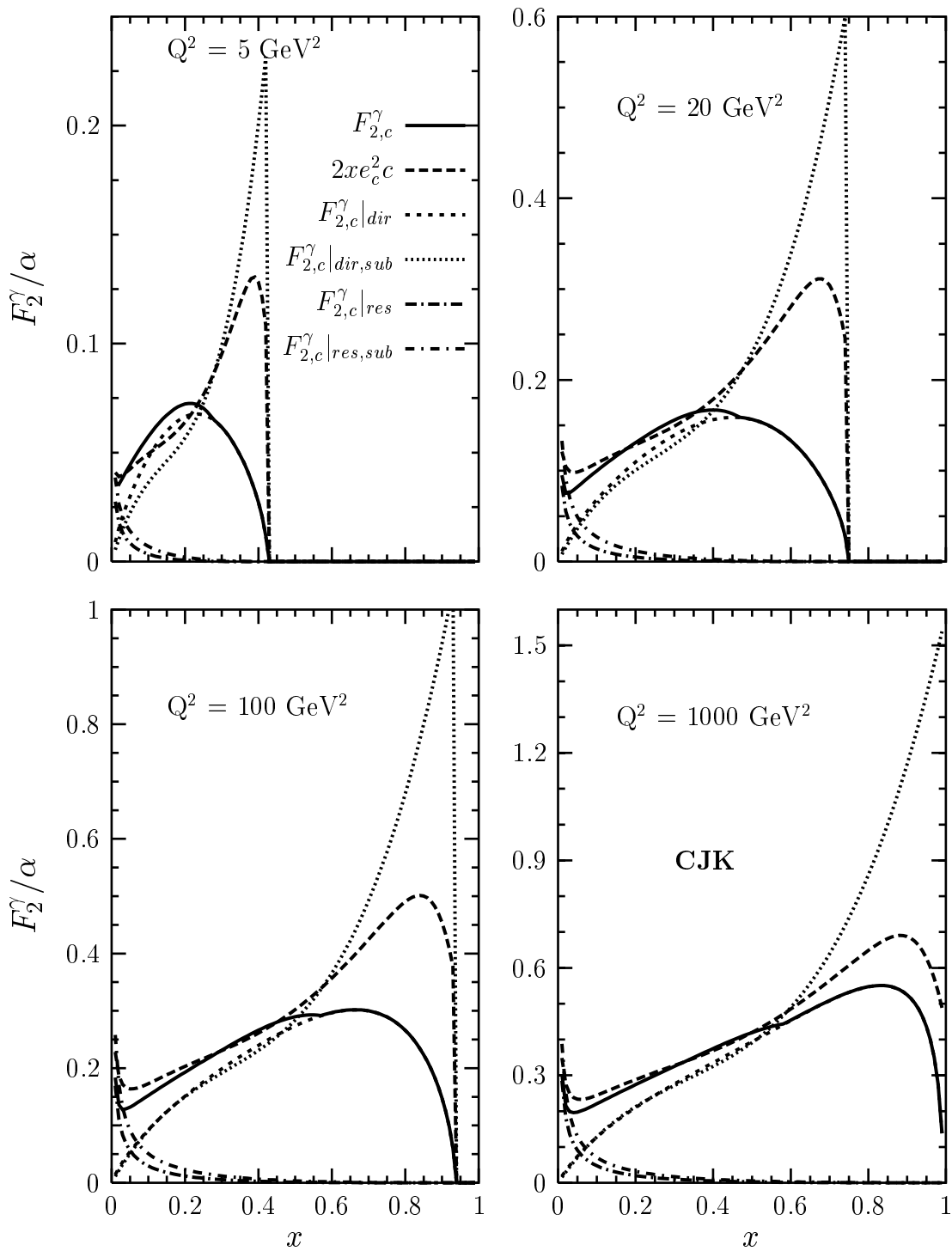}%
\caption{Comparison of various contributions to the photon structure
function $F_{2,c}^{\gamma}(x,Q^2)/\alpha$ in the CJK model for 
$Q^2=5,20,100$ and 1000 GeV$^2$.}
\label{acot}
\end{figure}

\clearpage

\begin{figure}[h]
\includegraphics[scale=1.0]{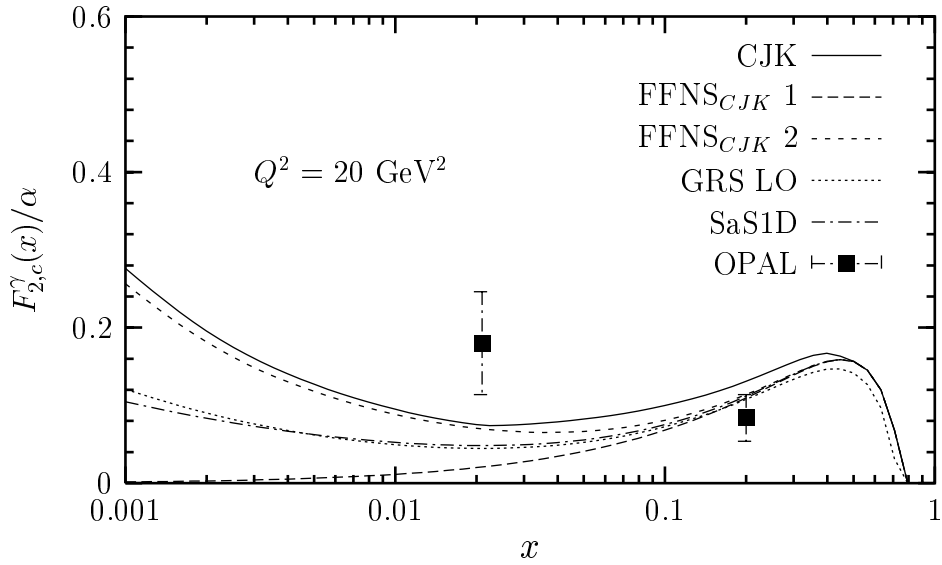}
\caption{Comparison of the structure function $F_{2,c}^{\gamma}(x,Q^2)/\alpha$
calculated in the CJK and FFNS$_{CJK}$ models and in GRS LO \cite{grs} and 
SaS1D \cite{sas} parametrizations with the OPAL measurement \cite{F2c}.} 
\label{fF2c}
\end{figure}

\begin{figure}[h]
\includegraphics[scale=1.0]{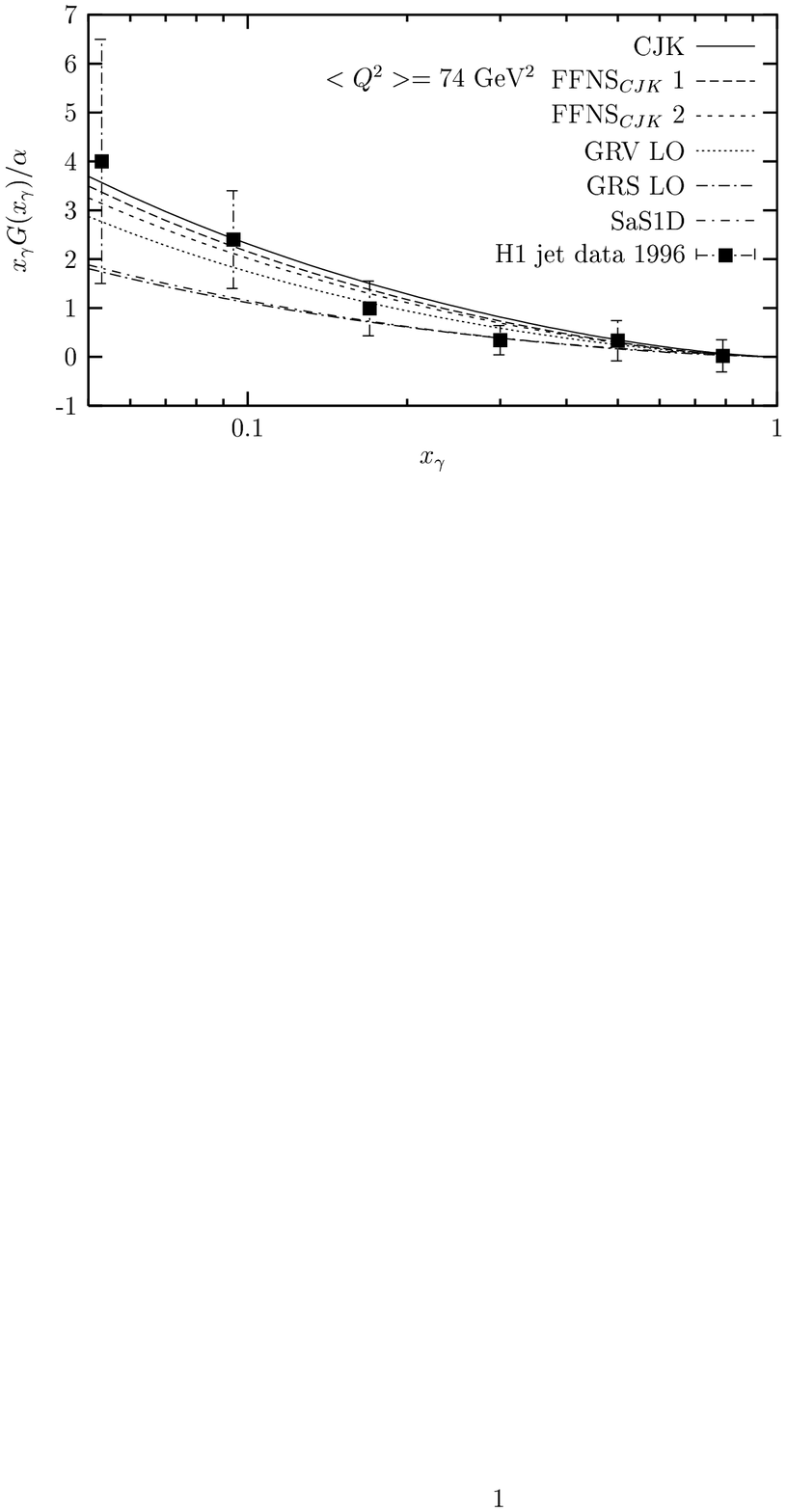}
\caption{Comparison of the gluon distribution obtained in the H1 measurement 
performed at $Q^2=74$ GeV$^2$ \cite{h1glu} with the predictions of the CJK, 
FFNS$_{CJK}$1 \& 2 models and GRV LO \cite{grv92}, GRS LO \cite{grs} and SaS1D 
\cite{sas} parametrizations with the OPAL measurement \cite{F2c}.} 
\label{gluh1}
\end{figure}

\clearpage

\begin{figure}[h]
\includegraphics[scale=1.0]{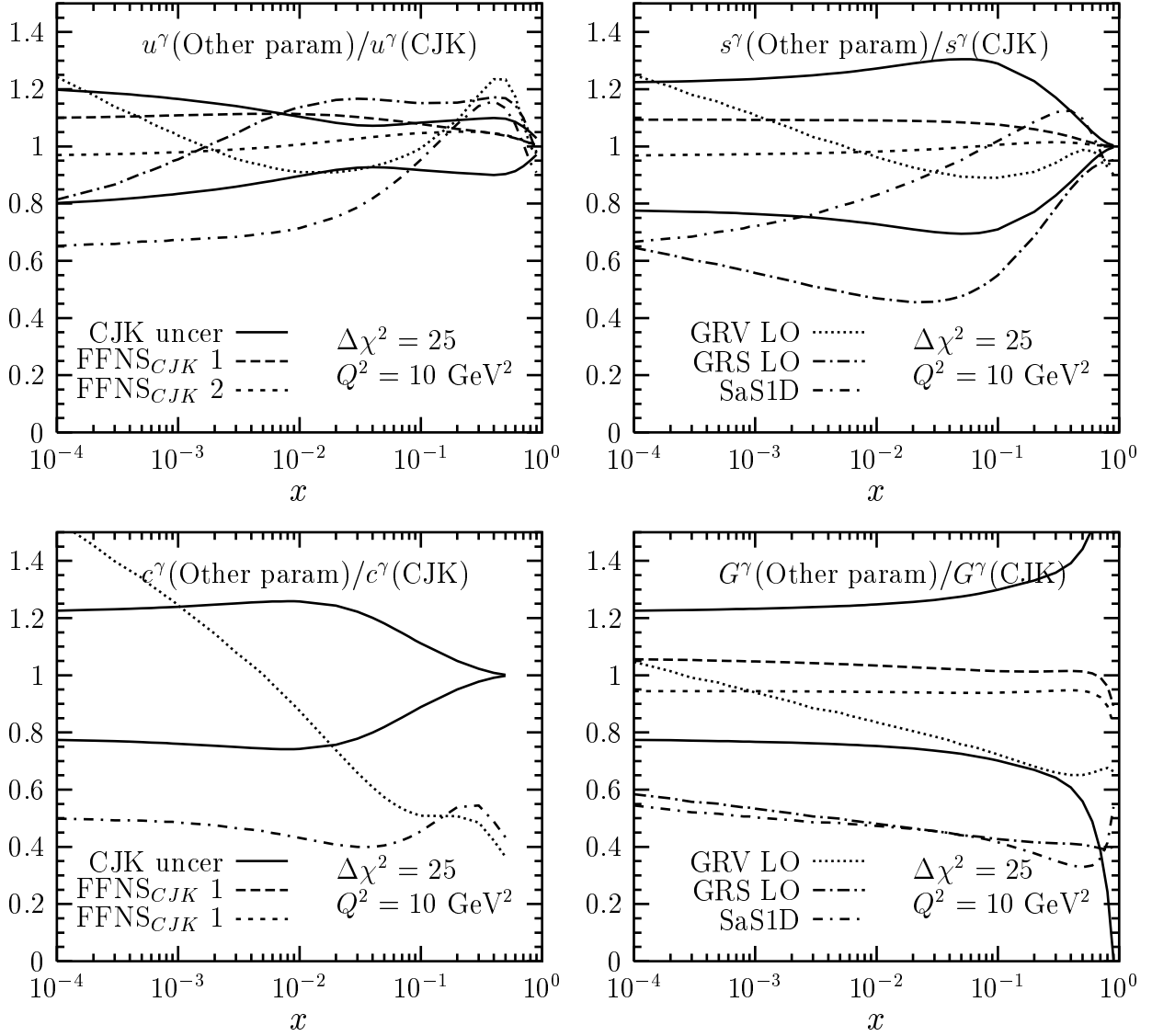}
\caption{Parton densities calculated in FFNS$_{CJK}$ models and
GRV LO \cite{grv92}, GRS LO \cite{grs} and SaS1D \cite{sas} parametrizations
compared with the CJK predictions. We plot for $Q^2=10$ GeV$^2$ the 
$q\gam(\mathrm{Other \: parametrization})/q\gam(\mathrm{CJK})$ ratios of the 
parton density calculated in the CJK model and its values obtained with 
other models and parametrizations. Solid lines show the CJK fit uncertainties 
for $\Delta \chi^2 = 25$ computed with the set of the $\{S_k^{\pm}\}$ test
parametrizations.}
\label{densuncup}
\end{figure}